\documentclass[aps,prl,twocolumn,showpacs,superscriptaddress,preprintnumbers,amsmath,amssymb]{revtex4-1}

\usepackage{graphicx}
\usepackage{epsfig}
\usepackage{dcolumn}
\usepackage{bm}
\usepackage{longtable}
\usepackage{color}

\begin{document}


\title{High superconducting anisotropy and weak vortex pinning in Co doped LaFeAsO}



\author{G.\ Li}
\affiliation{National High Magnetic Field Laboratory, Florida
State University, Tallahassee-FL 32310, USA}
\author{G.\ Grissonnanche}
\affiliation{National High Magnetic Field Laboratory, Florida
State University, Tallahassee-FL 32310, USA}
\author{J.-Q. Yan} \altaffiliation[Present address: ]{Materials Science and Technology Division, Oak Ridge National Laboratory, Oak Ridge, Tennessee 37831, and Department of Materials Science and Engineering, University of Tennessee, Knoxville, Tennessee 37996.}
\affiliation{Division of Materials Science and Engineering, Ames Laboratory, US-DOE, Iowa State University, Ames, Iowa 50011, USA}
\author{R W McCallum}
\affiliation{Division of Materials Science and Engineering, Ames Laboratory, US-DOE, Iowa State University, Ames, Iowa 50011, USA}
\author{T.A. Lograsso}
\affiliation{Division of Materials Science and Engineering, Ames Laboratory, US-DOE, Iowa State University, Ames, Iowa 50011, USA}
\author{H. D. Zhou}
\affiliation{National High Magnetic Field Laboratory, Florida
State University, Tallahassee-FL 32310, USA}
\author{L.\ Balicas} \email{balicas@magnet.fsu.edu}
\affiliation{National High Magnetic Field Laboratory, Florida
State University, Tallahassee-FL 32310, USA}


\date{\today}

\begin{abstract}
Here, we present an electrical transport study in single crystals of LaFe$_{0.92}$Co$_{0.08}$AsO ($T_c \simeq 9.1$ K) under high magnetic fields.
In contrast to most of the previously reported Fe based superconductors, and despite its relatively low $T_c$, LaFe$_{1-x}$Co$_x$AsO shows a superconducting anisotropy which is comparable to those seen for instance in the cuprates or $\gamma_H = H_{c2}^{ab}/H_{c2}^{c} = m_c/m_{ab} \simeq 9$, where $m_c/m_{ab}$ is the effective mass anisotropy. Although, in the present case and as in all Fe based superconductors, $\gamma \rightarrow 1$ as $T \rightarrow 0$. Under the application of an external field, we also observe a remarkable broadening of the superconducting transition particularly for fields applied along the inter-planar direction. Both observations indicate that the low dimensionality of LaFe$_{1-x}$Co$_x$AsO is likely to lead to a more complex vortex phase-diagram when compared to the other Fe arsenides and consequently, to a pronounced dissipation associated with the movement of vortices in a possible vortex liquid phase. When compared to, for instance, F-doped compounds pertaining to same family, we obtain rather small activation energies for the motion of vortices. This suggests that the disorder introduced by doping LaFeAsO with F is more effective in pinning the vortices than alloying it with Co.

\end{abstract}


\maketitle

\section{Introduction}

LaFePO was the first Fe-based pnictide compound to display a superconducting ground state at a transition temperature $T_c \simeq 7$ K. \cite{kamihara}. Soon after this discovery, Kamihara \emph{et al.} \cite{kamihara2} found the emergence of superconductivity, with a maximum $T_c$ of 26 K by doping its isostructural compound LaFeAsO with F. The first reported phase-diagram \cite{kamihara2}, comprises an antiferromagnetic metallic ground state that is progressively suppressed by F doping, which is found to produce a superconducting dome as previously observed in the cuprates (as a function of hole-doping). The boundary between antiferromagnetic and superconducting states suggests the coexistence between both phases although a subsequent phase-diagrams as a function of F content derived from either muon scattering \cite{luetkens} or thermal expansion measurements \cite{lwang} in polycrystalline material, indicates what seemingly is a first-order phase boundary between antiferromagnetic and superconducting phases with virtually no overlap between both states.

Soon after its discovery, the superconducting state in LaFeAsO$_{1-x}$F$_x$ was recognized to be unconventional. The experimental evidence includes, i) the absence of a coherence peak and the observation of a power law in the nuclear magnetic resonance (NMR) relaxation rate within the superconducting state \cite{nakai}, ii) the ratio of the superconducting transition temperature $T_c$ to the superfluid density is close to the so-called Uemura line for the high-$T_c$ cuprates \cite{luetkens2}, iii) an unconventional phase-boundary between superconducting and metallic states under high magnetic fields claimed to result from a multigap superconducting state \cite{hunte}, iv) the presence of pronounced antiferromagnetic spin fluctuations at temperatures above the superconducting transition temperature $(T_c)$ whose strength ``tracks" $T_c$  \cite{oka}, and v) the existence of a pseudogap- like phase preceding superconductivity \cite{ishida,boris,kondrat}. Electronic anisotropy, proximity to antiferromagnetism, pronounced antiferromagnetic spin-fluctuations within the superconducting and metallic phases, and the existence of a pseudogap state whose relation with the superconductivity is poorly understood, are all known properties of the high-$T_c$ cuprates \cite{palee}.

Nevertheless, a characteristic feature of the cuprates, is the broadening of the resistive transition in the presence of a magnetic
field. Initial investigations of the resistive transition in their so-called mixed state \cite{palstra}, indicated a current-independent, thermally activated behavior i.e. $\rho \sim \rho_0 \exp(-U_0/T)$, with $U_0$ ranging from $10^4$ K at high magnetic fields (for $H \simeq 10$ K) to $10^5$ K (for $H \simeq 10$ K). Transport studies \cite{koch, worthington} also revealed a characteristic temperature $T_g$ above which the current ($I$)-voltage ($V$) characteristics is linear, but becomes extremely non-linear below $T_g$: $V \propto \exp(-A/I^{\alpha})$. This observation was attributed to a transition between an unpinned viscous regime, or vortex-liquid to a pinned regime i.e. a vortex glass state, characterized by a limited motion of vortex lines. $T_g$ has been found to coincide with the irreversibility line as extracted from magnetometry measurements \cite{worthington}. In the vortex liquid regime and at lower temperatures the resistivity $\rho$ was found to display an activated behavior: $\rho \propto \exp(-U_0/T)$ with $U_0(T_k)\gg T_k$.

Here, we report electrical transport measurements in Co doped LaFeAsO samples in order to extract the phase-boundary between superconducting and metallic states as a function of magnetic field ($H$) and temperature ($T$). The resulting phase-diagram reveals a marked superconducting anisotropy $\gamma^{1/2} = H_{c2}^{ab}/H_{c2}^{c} = (m_c/m_{ab})^{1/2} \simeq 9$ in the neighborhood of $T_c$. This value for $\gamma$ is considerably larger than the values reported for other Fe based superconductors such as Ba$_{1-x}$K$_x$Fe$_2$As$_2$ \cite{singleton}, NdFeAsO$_{0.7}$F$_{0.3}$ \cite{jan}, K$_{0.8}$Fe$_{1.76}$Se$_2$ \cite{mun1}, and even Ca$_{10}$(Pt$_4$As$_8$)((Fe$_{1-x}$Pt$_x$)$_2$As$_2$)$_5$ \cite{mun2}. Although, as found in most Fe pnictides, $\gamma$ progressively tends to a value close to 1 as $T \rightarrow 0$ K, suggesting that the Pauli limiting mechanism, in contrast to the orbital effect, becomes the dominant pair breaking mechanism at low temperatures. Similarly to what is known from the family of cuprate superconductors, one observes a pronounced increase in the width of the resistive transition as a function of $T$ particularly for fields applied along the c-axis. An Arrhenius plot of the resistance as a function of temperature leads to extremely small values for the activation energy $U_0$ for vortex motion, i.e. between $10^1$ and $10^2$ K. Values in the order of $10^2$ K are obtained for fields applied along the ab-plane. These values contrast markedly with those measured in NdFeAsO$_{0.7}$F$_{0.3}$ \cite{jan} and with the large critical currents obtained in SmFeAsO$_{1-x}$F$_x$ single-crystals \cite{moll}, indicating that Co is far less effective than F in pinning vortices.

\section{Experimental}
LaFeAsO single crystals were synthesized in a NaAs flux at ambient pressure as described in Ref. \cite{yan}.
The quality of the so-obtained crystals was previously characterized by Laue backscattering, x-ray powder-diffraction, magnetization, and
resistivity measurements. The Co content was determined by using wavelength dispersive x-ray spectroscopy (WDS) in a JEOL JXA-8200 Superprobe electron probe microanalyzer (EPMA). Resistivity measurements under field were performed by using a conventional 4 terminal AC technique in either a Physical Parameter Measurement System or a Bitter resistive magnet, coupled to variable temperature insert, which is capable of reaching a field of 35 T.

\begin{figure}[htb]
\begin{center}
\epsfig{file=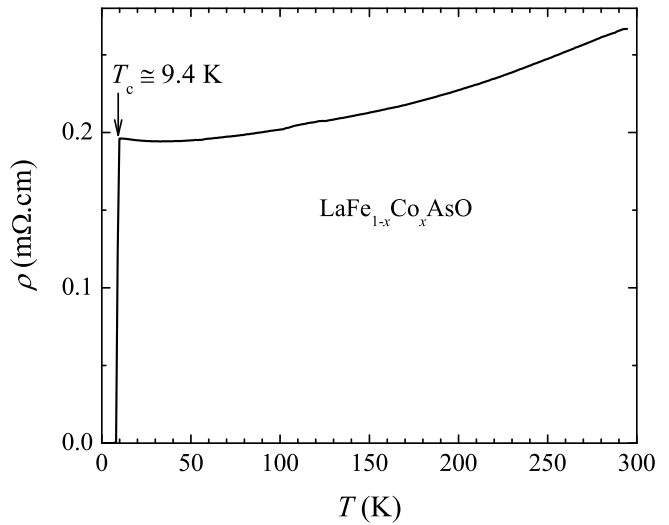, width = 8.6 cm}
\caption{Resistivity $\rho$ as a function of temperature $T$ for a LaFe$_{0.92}$Co$_{0.08}$AsO single crystal. The superconducting transition temperature $T_c$ is indicated by the vertical arrow. No evidence for either a structural or a magnetic phase-transition is observed.}
\end{center}
\end{figure}
Figure 1 shows the resistivity $\rho$ for LaFe$_{0.92}$Co$_{0.8}$AsO single crystal, for current flowing within the planes as a function of temperature and in absence of an externally applied field. One observes no clear indications for a phase transition, such as the orthorhombic distortion, or the antiferromagnetism seen in the parent compound \cite{kamihara, kamihara2,luetkens}. The resistivity ratio in the metallic state $\rho(300) \text{K}/ \rho(10 \text{K}) < 2$ is rather small suggesting that alloying with Co produces a considerable amount of disorder and it is an effective source of  quasiparticle scattering. For this level of Co doping the onset of the resistive transition is seen at $T_c \simeq 9.4$ K, with a transition width $\Delta T_c = T_c(90 \% \rho_n) -T_c(10 \% \rho_n) \simeq 1$ K, where $\rho_n$ is the resistivity in the metallic state just above the transition.

\begin{figure}[htb]
\begin{center}
\epsfig{file=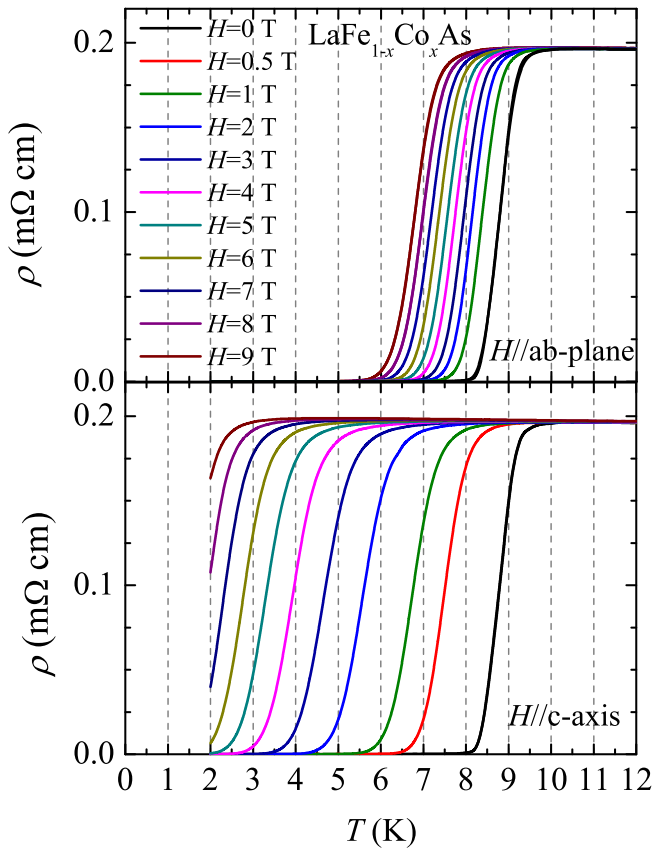, width = 8.6 cm}
\caption{(color online) Top panel: Resistivity $\rho$ as a function of temperature $T$ for a LaFe$_{0.92}$Co$_{0.08}$AsO single crystal measured under several values of the magnetic field $H$ applied along a planar direction.
Bottom panel: Same as in the top panel but for fields applied along the c-axis. }
\end{center}
\end{figure}
Figure 2 shows the resistive transition for a LaFe$_{0.92}$Co$_{0.08}$AsO single crystal as a function of the temperature for several values of the magnetic field up to 9 T, applied either along the ab-plane (top panel) or along the c-axis (bottom panel). While 9 T only suppresses $T_c$ by approximately 2 K for fields along the ab-plane, for fields along the c-axis the superconducting transition is seen to shift considerably to lower temperatures and under a field of 9 T, the transition temperature has shifted to temperatures below 2 K. For fields along either direction,  one observes what seemingly are parallel resistive transition curves, as usually seen in conventional superconductors, whose displacement in temperature is strongly orientation dependent.
\begin{figure}[htb]
\begin{center}
\epsfig{file=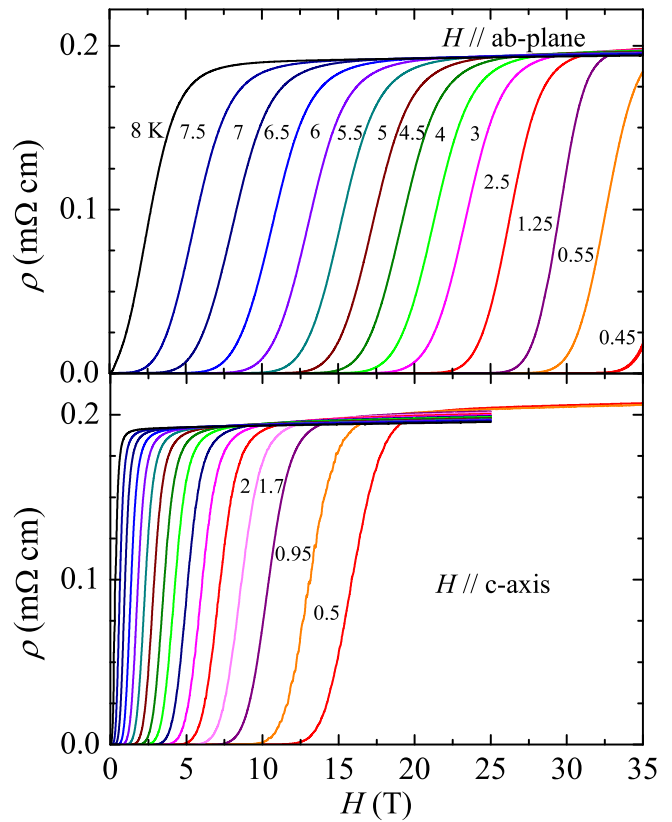, width = 8.6 cm}
\caption{(color online) Top panel: $\rho$ as a function of the magnetic field for a LaFe$_{0.92}$Co$_{0.8}$AsO single crystal and for several values of the temperature $T$.
Bottom panel: Same as in the top panel but for fields applied along the c-axis. }
\end{center}
\end{figure}

In order to construct the superconducting phase diagram LaFe$_{0.92}$Co$_{0.08}$AsO, as seen in Fig. 3, we measured the resistive transition at several temperatures and as a function of the applied field oriented either along the ab-plane (top panel) or along the c-axis (lower panel). As seen in Fig. 3, for fields aligned along the ab-plane, the resistive transition is just displaced to higher fields as $T$ is lowered, producing a set of nearly parallel resistive transition curves. But for fields oriented along the c-axis the width of the resistive transition is seen to increase as the $T$ is lowered. At first glance this would seem to be surprising since fluctuations should become less prominent as the temperature is lowered, and therefore one would naively expect the transition to sharpen as it is effectively seen for instance, in FeTe$_{1-x}$Se$_x$ \cite{tesfay}. However, such a broadening is commonly observed in the regime of thermally activated flux flow of vortices \cite{bardeen} which leads to a linearly dependent flux-flow resistivity behaving as $\rho_{\text{flow}} \simeq \rho_n B/H_{c2}$, i.e. the larger the upper critical field, or the lower the temperature, the smaller is the slope $B/H_{c2}$ as seen by us. $\rho_n$ is again the resistivity in the metallic state preceding the transition which, as previously stated, displays a weak field- and temperature-dependence. This would indicate that the energy barriers $U_0$ for the flow of vortices are effectively lower than the temperature at which the resistivity is measured \cite{vinokur}.

\begin{figure}[htb]
\begin{center}
\epsfig{file=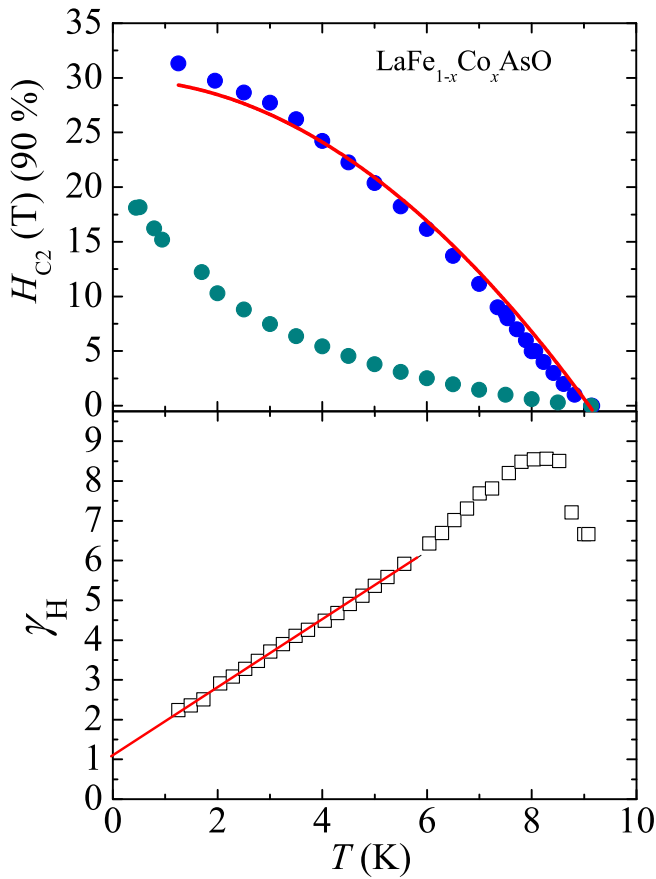, width = 8.6 cm}
\caption{(color online) Top panel: Magnetic field as a function of the temperature superconducting to metallic phase-boundary for a LaFe$_{0.92}$Co$_{0.08}$AsO single crystal, and respectively for fields applied
along the inter-planar direction (magenta markers) and along a planar direction (blue markers). To determine the phase boundary, we used the 50 \% of the value of the resistance in the normal state as the criteria (see text).
Lower panel: Superconducting anisotropy $\gamma_H = H_{c2}^{ab}/ H_{c2}^{c}$ as a function of temperature. Red line is a linear fit which extrapolates to $\gamma_H \simeq 1 $ as $T \rightarrow 0$.}
\end{center}
\end{figure}

The resulting superconducting phase diagram is shown in Fig. 4, where we  used the middle point of the resistive transition, i.e. $T_c^m = T (0.5 \rho_n)$, with $\rho_n$ being the resistivity in the metallic state preceding the temperature dependent resistive transition, or the value where $\rho_n$ starts to deviate from the behavior displayed by the metallic state magneto-resistivity ($\rho_n(H)$). $\rho_n(H)$ was adjusted to a second-degree polynomial. As seen in Fig. 4 the upper critical field for fields applied along the ab-plane $H_{c2}^{ab}$ follows a concave down curvature which extrapolates to $H_{c2}^{ab} (T = 0 \text{K}) = \phi_0/(2 \pi \xi_{ab} \xi_c) \sim 32.5$ K, which corresponds to $\xi_{ab} \xi_c \sim 1014 $ {\AA}$^2$ where $\xi_{ab}$ is the in-plane superconducting coherence length and $\xi_c$ is the inter-plane one. As for fields applied along the c-axis, one observes the usual concave-up curvature for $H_{c2}^{c}(T)$ claimed to result from multi-band superconductivity \cite{hunte}, and whose extrapolation to zero temperature seems to saturate at a value of $\sim 20$ T corresponding to $\xi_{ab} \sim 40.6$ {\AA} and therefore implying $\xi_{c} \sim 25$ {\AA} which is considerably larger than the inter-plane distance c $= 8.746$ {\AA} \cite{yan}. The red line is a fit of $H_{c2}^{ab} (T)$ to the conventional empirical expression:
\begin{center}
\begin{equation}
H_{c2}^{ab} (T) = H_{c2}^{ab} (0) \left( 1-(T/T_c)^2 \right)
\end{equation}
\end{center}

The fit, which is relatively poor at low temperatures, yields a lower value for $H_{c2}^{ab} (0) \simeq 29.9$ T corresponding to $\xi_{ab} \xi_c \sim 1102 $ {\AA}$^2$.
The deviations with respect to conventional behavior is rather intriguing and bears resemblance with a previous report in Fe$_{1+y}$Te$_{1-x}$Se$_x$ \cite{tesfay}, where an upturn is observed in $H_{c2}(T)$ at low temperatures and which was interpreted as evidence for an additional phase-transition, the so-called Fulde-Ferrell-Larkin-Ovchnnikov (FFLO) superconducting state \cite{fflo}. This deviation represents extremely weak evidence for an additional superconducting state. Although the upper critical field in this material is considerably larger than the weak coupling Pauli limiting field value $H_p = \Delta_0 /2 \mu_B$ with $\Delta_0 = 1.75 k_B T_c$ which leads to the standard expression $H_p /T_c =$ 1.84 T/K, i.e. $T_c \simeq 9.4$ K leads to $H_p \simeq 17.3$ T. This value is nearly a factor of 2 smaller than the extrapolation to zero temperature of the experimentally observed upper-critical field for fields along the ab-plane. This could be understood if the correlations were particularly strong in this system renormalizing $H_{c2}$ considerably, and suggesting that this system is indeed Pauli limited. Under such circumstances additional superconducting phases such as the FFLO state become a possibility.

The lower panel of Fig. 4 shows the resulting temperature dependence for the anisotropy in upper critical field $\gamma_H = H_{c2}^{ab}/H_{c2}^{c}$ which reflects the anisotropy in the effective band mass. It is observed to initially increase up to value of $\simeq 9$ close to $T_c$, which is nearly twice the value of 4 to 5 previously reported for a ``1111" compound \cite{jan}, but $\gamma$ quickly decreases with further decreasing the temperature and a simple linear extrapolation suggests an isotropic system in the limit of zero temperature. This temperature dependent behavior which is seen in virtually all Fe based superconductors, can be understood if one assumes that the orbital limiting field is the dominant pair-breaking effect at higher temperatures. But the Pauli liming field, which depends on the value of the superconducting gap and on the anisotropy of the Land\'{e} $g$-factor, becomes the dominant one at low temperatures (relative to $T_c$) if the $g$ factor is nearly isotropic. This would further indicatte that this system is Pauli limited. To date, and to our knowledge this is the superconducting phase-diagram extracted over the widest range in reduced temperature $t=T/T_c$ for a 1111 Fe arsenide compound.
\begin{figure}[htb]
\begin{center}
\epsfig{file=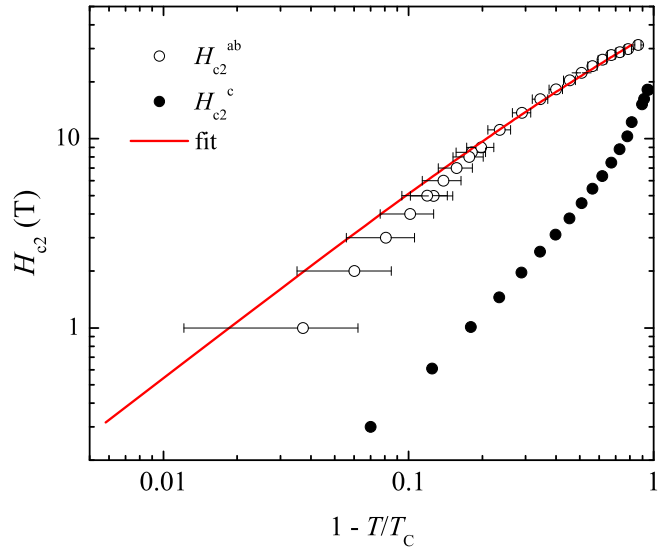, width = 8.6 cm }
\caption{(color online) Upper critical fields $H_{c2}^{c}$ (solid markers) and $H_{c2}^{ab}$ (open markers) as a function of $1-t$ where $t = T/T_c$. Red line is a fit of $H_{c2}^{ab}$ to Eq. (1) which yields a Pauli
limiting field $H_p^{\|ab} = (64 \pm 11)$ T, and an orbital limiting field $H_0^{\|ab} = (54.6 \pm 3.1)$ T corresponding to $\xi_c \xi_{ab} = (603 \pm 34)$ {\AA}$^2$.}
\end{center}
\end{figure}

In order to evaluate the contributions of both orbital and Pauli
pairbreaking effects, and in order to evaluate the so-called Maki parameter $\alpha_M =
\sqrt{2}H_\text{o}/H_p$, where $H_\text{o}$ is the orbital limiting field,
we analyze our $H_{c2}(T)$ data at temperatures close to $T_c$
where the Ginzburg-Landau theory yields Ref. \cite{Gurevich2010}:
    \begin{equation}
    \left( \frac{H}{H_p}\right)^2 + \frac{H}{H_\text{o}} = 1 - \frac{T}{T_c}
    \label{gl}
    \end{equation}
Very close to the critical temperature, $(T_c-T)/T_c \ll
(H_p/H_\text{o})^2$, the first paramagnetic term in the left hand
side is negligible and Eq. (\ref{gl}) yields the orbital linear GL
temperature dependence, $H_{c2}=H_\text{o}(1-T/T_c)$.  At lower
temperatures, $(T_c-T)/T_c > (H_p/H_\text{o})^2$, the Pauli limiting
field $H_p$ dominates the shape of $H_{c2}(T) \propto (1-t)^{1/2} $
even in the GL domain if $H_p < H_\text{o}$. The latter inequality
is equivalent to the condition that the Maki parameter $\alpha_M
\sim H_\text{o}/H_p >1$ is large enough, assuring that the
paramagnetic effects are essential.  Shown in Fig. 5 are the log-log
plot of our $H_{c2}(T)$ as a function of $1-T/T_c$ where the red
line is a fits to Eq. (\ref{gl}). Given the unconventional concave up curvature
for fields applied along the c-axis, this fit can only be applied to the data
where the field is oriented along the ab-plane. The fit is excellent for the
high-field region, but less so for temperatures close to $T_c$, probably due to the relatively large error bars in
determining the temperature ($\Delta T \sim 25$ mK) which are inherent
to transport measurements. It might also be attributed to the broadening of the
resistive transition due to local $T_c$ inhomogeneities.
The fit yields $H_p^{\|ab} = (64 \pm 11)$ T for the Pauli limiting field, and $H_0^{\|ab} = (54.63 \pm 3.1)$
T for the orbital limiting field. The fit yields values that are comparable in magnitude, making it difficult to distinguish
or evaluate the dominant pair breaking mechanism at low temperatures, therefore suggesting a Maki parameter close to unity, which is nearly beyond the validity of Eq. (2).
Defining the effective Ginzburg-Landau coherence lengths,
$\xi_{ab}(T)=\xi_{ab}(1-T/T_c)^{-1/2}$ and
$\xi_{c}(T)=\xi_{c}(1-T/T_c)^{-1/2}$, we obtain $\xi_c \xi_{ab} = (\phi_0/2\pi
H_{\text{o}}^{\|ab})= (603 \pm 34)${\AA}$^2$, which is considerably smaller than the value of 1014{\AA}$^2$ estimated from $H_{c2}^{ab}$.
\begin{figure}[htb]
\begin{center}
\epsfig{file=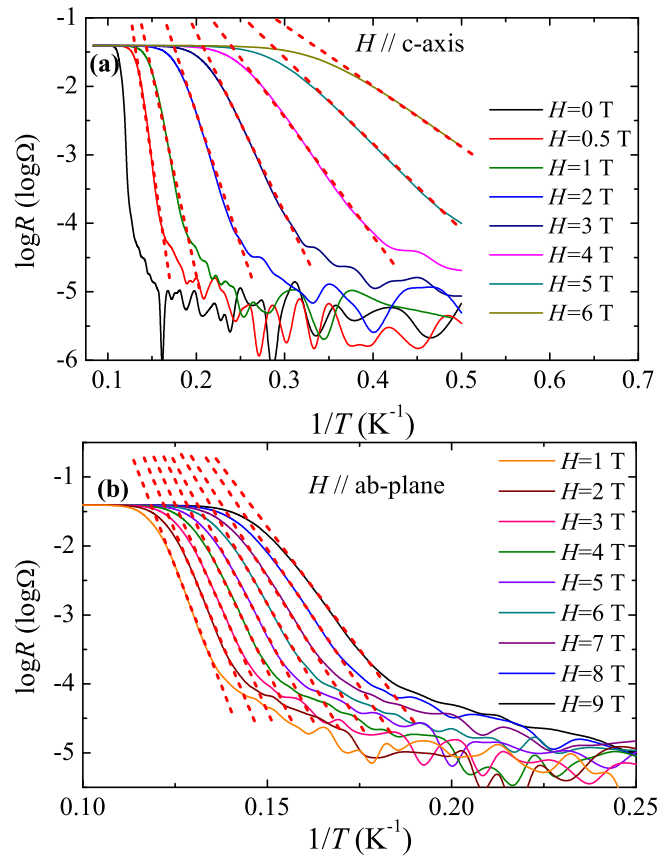, width = 8.6 cm}
\caption{(color online) (a) Logarithm of the resistance as a function of the inverse of the temperature $T^{-1}$ for fields applied along the c-axis. (b) Same as in (a) but for fields along the ab-plane. Dotted red lines, are guides to the eyes, indicating a region in $T^{-1}$ where the resistance across the superconducting transition clearly displays activated behavior.}
\end{center}
\end{figure}

Given the relatively large superconducting anisotropy observed here, comparable for instance to values reported for the least anisotropic cuprates, it is pertinent to ask if it would have any significant effect on the vortex phase diagram of this material. In effect, Figs. 6 (a) and 6 (b) show the logarithm of the resistance ($\log R$), as it decreases through the superconducting transition, and as a function of the inverse of the temperature $T^{-1}$ for several values of the magnetic field applied either along the c-axis or along an in-plane direction, respectively. As seen, for over two decades in temperature $\log R$ is linear in $T^{-1}$ allowing us to extract the field-dependence of the activation energy $U_0$. The zero-field curve does not display any clear linear dependence over a significant range in temperatures. The saturation observed at lower temperatures is mostly due to the noise floor of our instrumental set-up, and probably also to a crossover towards a pinned vortex regime, i.e. the vortex-solid state.
\begin{figure}[htb]
\begin{center}
\epsfig{file=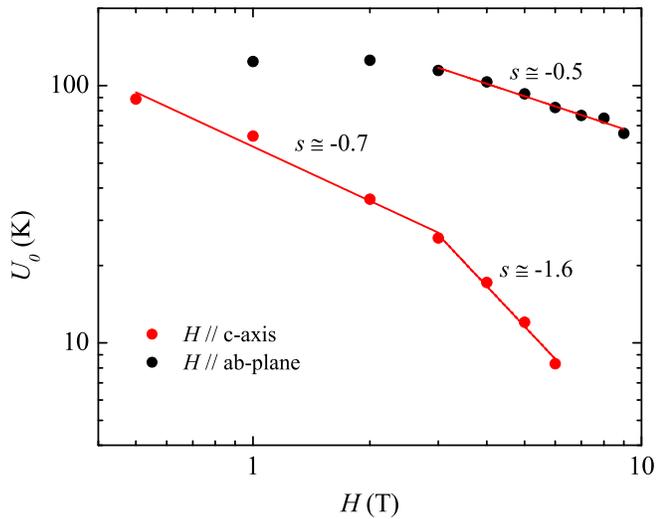, width = 8.6 cm}
\caption{(color online) Activation energies as a function of the magnetic field as extracted from the Arrhenius plots in Figs. 6 (a) and (b), respectively for  $H \parallel$ c-axis (red markers) and for $H \parallel$ ab-plane (black markers). Red lines are linear fits indicating the respective power laws.}
\end{center}
\end{figure}

Figure 7 shows the resulting field dependence for the activation energy $U_0$ for both orientations of the external field. What is remarkable in the present case, is the extremely small values of $U_0$ which are nearly 2 to 3 orders of magnitude smaller than the values reported for either the cuprates \cite{palstra} or the 1111 Fe-pnictides \cite{jan}. This result is particularly surprising, since Co is incorporated within the superconducting Fe arsenide planes, and therefore one would naively expect it to act as a quite effective point pinning center for vortices. This result is particularly difficult to understand if one considers that F$^{-}$ is incorporated within the nearly electronically inert rare-earth oxide layer having an ionic radius in the order of $\sim$ 1 {\AA}, thus being just about 15 \% larger then the ionic radius of Co$^{+2}$ which in addition, is expected to be nearly magnetic. Another surprising result is the large anisotropy (nearly one order of magnitude) between the high field values of $U_0$ for fields applied either perpendicularly or along the conducting planes. This suggests that the layered structure of this material is more effective in pinning vortices than the incorporation of about 8 \% of point disorder. At the moment, these observations suggest rather unconventional vortex pinning mechanisms for the Fe arsenide superconductors. This figure also shows the power-law dependence in field or $U_0 \propto H^{s}$, where for fields along the ab-plane $U_0$ remains nearly constant followed by a rather weak power law with $s = -0.5$. For fields along the c-axis on the other hand, one observes a weak power law, i.e. $s= -0.7$ which crossovers to a value $s = -1.6$ when $H>3$ T suggesting collective creep at high fields \cite{gianni}.

\section{Summary}

In summary, this first single-crystal electrical transport study on a La based ``1111" Fe-arsenide compound reveals a relatively large superconducting anisotropy, i.e. nearly two times larger than the anisotropy previously reported in a Nd-based 1111 compound \cite{jan}, suggesting perhaps that a larger electronic anisotropy is in effect detrimental to the superconducting transition temperature in these compounds. Perhaps, not surprising, anisotropies on the order 9, combined with relatively weak vortex pinning by point defects, lead to behavior akin to what is seen in the vortex-liquid phase of the cuprates \cite{vinokur}. However, the extremely small activation energies for vortex flow, as extracted here, indicate that the introduction of point defects in the FeAs planes is ineffective in pinning vortices. This is extremely difficult to understand when compared to the strong pinning reported for F doped samples \cite{jan, moll} or for the Co doped 122 compounds \cite{prozorov} and will require major experimental and theoretical efforts to elucidate such a contrast.

\section{Acknowledgements}
The NHMFL is supported by NSF through NSF-DMR-0084173 and the
State of Florida.  L.~B. is supported by DOE-BES through award DE-SC0002613.
JW, TG and TS acknowledge support from FSU. Work done at the Ames Laboratory was supported by the U.S. Department of Energy, Office of Basic Energy Sciences, Division of Materials Sciences and Engineering.  Ames Laboratory, which is operated for the U.S. Department of Energy by Iowa State University under Contract No. DE-AC02-07CH11358.

\end{document}